\documentstyle[twoside,fleqn,espcrc2]{article}

\newcommand{\AmS}{{\protect\the\textfont2
  A\kern-.1667em\lower.5ex\hbox{M}\kern-.125emS}}

\newcommand{\beq}{\begin{equation}}
\newcommand{\eeq}{\end{equation}}
\newcommand{\ber}{\begin{eqnarray}}
\newcommand{\eer}{\end{eqnarray}}

\title{Effective Theories for Exclusive and Seminclusive Processes and 
       Factorization}

\author{ U. Aglietti\address{Dipartimento di Fisica, Universita' di Roma
         `La Sapienza', \\
         INFN, Sezione di Roma I, P.le A. Moro 2, 00185 Roma, Italy }%
\thanks{ Work done in collaboration with Dr. Corbo' in refs.[1,2] }
}

\begin{document}

\begin{abstract}

The effective theories for massless quarks describing
exclusive and seminclusive processes are discussed,
considering in particular the factorization problem.

\end{abstract}

\maketitle

\section{Introduction}
\label{sec1}

Factorization is the property that a hadronic matrix element
can be written as the product of simpler matrix elements.
For example, the decay
\beq\label{eq1}
B~\rightarrow~D+\pi
\eeq                              
has amplitudes which can be written, if factorization
holds, as
\ber
&&g^{\mu\nu}~
\langle D,\pi\mid J_{\mu,i}^{b\rightarrow c} J_{\nu,i}^{u\rightarrow d}
\mid B\rangle
\nonumber\\ \label{eq2}
 &=& g^{\mu\nu}~\langle D \mid  J_{\mu,i}^{b\rightarrow c}\mid B\rangle ~
\langle \pi\mid J_{\nu,i}^{u\rightarrow d}\mid 0\rangle
\eer
where $J_i^{\mu}$ is the color singlet (octet) weak current
for $i=1(8)$
\footnote{We do not consider the problem of the different
renormalization scale dependence 
of the two sides of eq.(\ref{eq2}). For the scale choice
problem see for example ref.\cite{cit3}. }.
The above property implies that the rate of (\ref{eq1})
can be expressed 
in terms of the rate of the semileptonic decay
\beq\label{eq3}
B~\rightarrow~D+l+\nu,
\eeq
whose amplitude is proportional to the matrix element
\beq\label{eq4}
\langle D \mid J_{\mu,1}^{b\rightarrow c}\mid B\rangle,
\eeq
and of the (purely) leptonic decay rate of
\beq\label{eq5}
\pi~\rightarrow~ l + \nu,
\eeq                       
whose amplitude is proportional to
\footnote{The matrix elements on the right hand side of eq.(\ref{eq2})
vanish for $i=8$ because of color conservation.}
\beq\label{eq6}
\langle 0\mid J_{\nu,1}^{u\rightarrow d}\mid \pi\rangle
~=~i f_{\pi} p_{\nu}.
\eeq
Factorization in general allows to relate rates of nonleptonic decays
to the rates of simpler processes, involving fewer hadrons
in the external states. 
It is often assumed as a reasonable hypothesis to
analyse experimental data \cite{cit4}or to make phenomenological
predictions, for example in $CP$-violating $B$-processes.
It is notoriously difficult to prove factorization in the framework
of $QCD$, so that a different approach is to identify the 
relevant degrees of freedom in the process and replace $QCD$
with a simpler, effective theory.

The starting point of our discussion is a physical picture
of the decay (\ref{eq1}) due to Bjorken \cite{cit5},
which strongly suggests factorization.   
In the limit of a very large beauty mass $m_b$, 
the $\overline{u}$ and $d$ quarks from the $b$-fragmentation
are emitted in a color singlet state 
\footnote{See later for a discussion of the color octet state.}
with a very large energy
$\sim m_b$.
According to a classical picture, they are created in the same
spatial point and separate very slowly from each other because
of time dilation. This implies that the color dipole is zero
at the decay time and remains very small during the evolution, 
when the quarks leave the cloud region of the $B$-meson;
hadronization occurs at a much later stage.
The interaction of the light pair with the heavy meson
is negligible, so the two subsystems evolve
independently and  factorization results.    

The above idea is realized in quantum field theory taking
the limit
\beq\label{eq7}
E~\rightarrow~\infty,
\eeq
where $E$ is the energy of a massless quark.

In sec.\ref{sec2} we sketch the resulting   
`Large Energy effective Theory' $(LEET)$ \cite{cit6}.  
The main property of the limit (\ref{eq7}) is that 
quantum fluctuations in the motion are suppressed and the 
quark `resembles' a classical particle, in  
analogy with the limit of an infinite mass
in the HQET \cite{cit7}.
In particular, the transverse motion of the quark with respect to the line
of flight (the direction $n$) is completely neglected.
This implies that quarks created in the same point 
do not separate from each other {\it at any time}.
There is not any dipole color grow (as it does occur in Bjorken picture) 
to account for hadronization into a $\pi$.
As we will see explicitly, the limit (\ref{eq7}) is `too strong' to describe 
an exclusive process.

In sec.\ref{sec3} we show that the $LEET$ can describe
seminclusive decays, such as for example
\beq\label{eq8}
B~\rightarrow~D+jet
\eeq
and can be used to prove factorization for this class
of processes.

In sec.\ref{sec4} we come back to the problem of describing
an exclusive process such as (\ref{eq1}) in the effective theory
framework and we introduce a new effective theory containing
the required transverse momentum terms.

Sec.\ref{sec5} contains the conclusions and an outlook
to future developments.
 
\section{ The LEET and its failure to describe exclusive processes  }
\label{sec2}

Let us briefly review the derivation of the $LEET$ propagator 
\cite{cit6,cit7}.
The momentum $P$ of a massless quark is decomposed 
into a `classical part' $E n$ and a fluctuation $k$:
\beq\label{eq9}
P~=~En+k.
\eeq
$n$ is a light-like vector, $n^2=0$, normalized by the condition
$v\cdot n=1$, where $v$ is a reference time-like vector, $v^2=1$,
with positive time component, $v_0>0$.
$E$ is the classical, `primordial', energy of the quark
in the rest frame of $v$ and has to be considered large,
\beq\label{eq10}
\mid k_{\mu} \mid~\ll~E.
\eeq
The propagator is given by
\ber
iS_0(En+k)&=&i\frac{E\hat{n}+\hat{k}}{(E n+k)^2+i\epsilon} 
\nonumber\\
&=& i\frac{\hat{n}/2+\hat{k}/(2 E)}{ n\cdot k + k^2/(2 E) + i\epsilon} 
\nonumber\\ 
\label{eq11}
&\simeq&\frac{\hat{n}}{2}~\frac{i}{n\cdot k + i \epsilon},
\eer
where in the last line the limit (\ref{eq7}) has been
taken. The energy-momentum relation is
\beq\label{eq12}
\epsilon~=~\vec{u}\cdot \vec{k}~=~k_z,
\eeq
where $\vec{u}\doteq\vec{n}/n_0$ is the kinematical velocity, with
$\mid\vec{u}\mid=1$.
In the last term a motion along the $z$-axis has been assumed.

\noindent
The propagator in configuration space is the scalar density
of a particle moving along a ray with the velocity of light: 
\beq\label{eq13}
iS_0(x)~=~\frac{\hat{n}}{2}~\theta(t)~
\frac{\delta^{(3)}(\vec{x}-\vec{u}t)}{n_0}.
\eeq
The interaction with the gauge field produces a P-line factor
joining the origin with the point $x$ along the light-like
trajectory specified by $n$:
\beq\label{eq14}
iS(x)=iS_0(x)P\exp\Bigg[ig\int_0^{t/n_0}ds n_{\mu}A^{\mu}(n s)\Bigg].
\label{prop}\eeq
Note the factorization of both spin and color degrees of
freedom.
As it is clear from the derivation, 
the $LEET$ propagator describes massless particles suffering 
soft interactions only. In other words,
the effect of the quark on the gluon field is fully taken into account,
while the reaction of the field on the particle is completely
neglected. The limit (\ref{eq7}) is a systematic
no recoil approximation and is called 
(massless) eikonal model in perturbative $QCD$ \cite{cit7b}. 

We present now a physical argument of the failure of the $LEET$
to describe an exclusive decay such as (\ref{eq1}).
Since the $\overline{u}$ quark is approximated by a $LEET$ quark, 
the effect of the $d$ on the $\overline{u}$ is completely neglected. 
Analogously, since the $d$ quark also is taken in the $LEET$, 
the effect of the $\overline{u}$ on the $d$ is neglected. 
As a consequence, there
is not any dynamics of the `valence' quarks, and consequently
not a  chance of bounding them together into a $\pi$.
This conclusion is clear also from the form (\ref{eq13}) 
of the propagators,
according to which quarks move on a prescribed trajectory.  

The irrelevance of the $LEET$ in the exclusive domain
can be analytically confirmed with a spectral decomposition
of the correlation functions associated to the decay (\ref{eq1})
or with a study of the
one-loop pinch singularities of Feynman diagrams \cite{cit1,cit2}.

\section{Seminclusive decays and factorization by means of the LEET}
\label{sec3}

The argument of the previous section is centered on the fact
that the $\overline{u}$ and $d$ quarks are required to combine into 
a single meson state in the decay (\ref{eq1}):
they are the relevant degrees of freedom of the system,
the meson valence quarks.
If we consider instead the seminclusive decay (\ref{eq8}),
many other degrees of freedom are involved, participating
to the jet development.
In the latter case, we can identify the $\overline{u}$ and $d$ quarks 
with partons, i.e. with high energy excitations,
emitting (softer) secondary quarks and gluons in the parton cascade.  
If we neglect the reaction of the secondary particles,
we can legitimately replace the $\overline{u}d$ pair
with $LEET$ quarks.
Assuming a jet with a small angular width
\footnote{As it is well known, a proper definition
of a jet involves also an energy resolution parameter
in order to cancel infrared singularities. 
We do not consider this problem explicitly, which deserves
further investigation. We wish to thank Prof. G. Altarelli
for having brought this point to our attention.},
we can take equal velocities for the quarks.
Apart from irrelevant constants,
the dynamics of the pair is represented, before
functional integration over the gluon field, by the
factor 
\beq\label{eq15}
Tr\Bigg[ Pe^{ig\int_0^x A_{\mu} dx^{\mu}} 
         Pe^{ig\int_x^o A_{\mu} dx^{\mu}} \xi_i\Bigg] =
Tr \Big[\xi_i\Big], 
\eeq
where the weak current is taken in the origin,
the light pair interpolating field is a color singlet bilinear
placed in $x$,
$\xi_i=1,t_a$ for $i=1,8$ respectively
and $Tr[\xi_i]=N_C \delta_{i,1}$.
The above factor is independent of the gluon field, 
implying the absence of any interaction of the `light'
system with the `heavy' system.
This implies factorization for
the seminclusive decay in the spirit of Bjorken
color screening argument.

\section{A new effective theory}
\label{sec4}

Let us consider a meson in the infinite momentum frame 
moving along the $z$-axis: that is equivalent to the limit 
(\ref{eq7}) discussed in sec.\ref{sec2}. 
We observe the valence quarks exchanging transverse momenta
of order of the hadronic scale
$\Lambda_{QCD}$: 
$P_T~\sim~\Lambda_{QCD}$\footnote{We neglect perturbative corrections
coming from hard gluons producing a broad tail in the quark
transverse momentum distribution of the form
$\alpha_S ~ d k_T^2/k_T^2$.}.
These interactions are necessary for the bound state dynamics, while
they are corrections of order $\Lambda_{QCD}/E$
to the leading term $P_z\sim E$, and are
neglected in the limit (\ref{eq7}). 
Therefore the failure of the $LEET$ to describe 
exclusive processes is of kinematical origin.
Eq.(\ref{eq12}) is the lowest order term in the expansion
of the energy-momentum relation $P^2=0$ with $P$ given in eq.(\ref{eq9}):
\ber
E + \epsilon &=& \sqrt{(E+k_z)^2 + k_T^2} 
\nonumber\\
\label{eq16}
&=& E + k_z + \frac{k_T^2}{2 E} + \ldots
\eer
We can `correct' the $LEET$ by including 
the term of order $1/E$  into the propagator:
\beq\label{eq17}
i S(k)~=~\frac{\hat{n}}{2}~
\frac{i}{n\cdot k-{\vec{k}_T}^2/2E~+i\epsilon}
\eeq
where $n^{\mu}=(1;0,0,1),~k_T^{\mu}=(0;\vec{k}_T,0)$

Eq.(\ref{eq17}) defines an effective theory:
the propagator contains the 4-velocity $n$ as an external vector,
so there is a formal breaking of Lorentz invariance as it happens
in the $LEET$ or in the $HQET$ \cite{cit8}.   
Furthermore $iS(k)$ is forward in time, so that antiparticles
are removed and the vacuum is consequently 
trivial. This is expected from an effective theory
describing hard partons, because 
particle-antiparticle pairs have a threshold energy
of order $2E$ and cannot be excited with soft interactions.
This new effective theory, which we called $\overline{LEET}$, 
is much more complicated than the $LEET$.  
In particular, the hard scale $E$
is still present in the theory, i.e. it cannot completely be
removed.

The propagator is given in configuration space by 
\beq\label{eq18}
iS(t,\vec{x})~=~
\frac{\hat{n}}{2}~
\theta(t)~\delta(z-t)~\frac{E}{2\pi it}~e^{iE b^2/(2t)}
\eeq
where $b=\mid\vec{x}_T\mid$ is the impact parameter.
The effect of the transverse momentum term is factorized
(compare with eq.(\ref{eq13})) and produces fast oscillations
of the wavefunction with $b$ at small times.
The larger the energy $E$, the faster the oscillations are
\footnote{In the limit $E\rightarrow\infty$ we recover 
eq.(\ref{eq13}) in the well known distribution sense.}.   
The transition from the $LEET$ to the $\overline{LEET}$
is analogous to the transition from geometrical
optics to physical optics, the latter taking into account diffraction
in first approximation.
It is interesting to note that 
after euclidean continuation, $(t \rightarrow -i t)$,
the last two terms in eq.(\ref{eq18}) represent 
a gaussian diffusion in the impact parameter space.

At $b=0$ we have:
\beq\label{eq19}
S(t,\vec{x})~\simeq~\frac{\hat{n}}{2}~
\theta(t)~\delta(z-t)~\frac{E}{2\pi it}.
\eeq
There is a diffusion normal to the classical particle
trajectory $z=t$ produced by transverse momentum fluctuations,
which is instead absent in the $LEET$.
The amplitude for the particle to remain into the classical
trajectory decays like $1/t$, so the probability decays like
$1/t^2$.

We believe that the $\overline{LEET}$ is the correct effective
theory for massless particles as long as exclusive processes
are concerned. 

It is hard to reach a conclusion about factorization in the
exclusive decay (\ref{eq1}) on the basis of simple analytical
computations with the $\overline{LEET}$.
That is because of diffusion in the impact parameter space, according
to which light quark dynamics is represented by a superposition
of non trivial Wilson loops instead of a single (trivial) one.
  
\section{Conclusions and Outlook}
\label{sec5}

The conclusions of our analysis consists of various
statements, both of negative and positive character,
which can be organized in the following way.

\noindent
$i)$
Exclusive decays such as (\ref{eq1}) cannot be described by the $LEET$,
which is a too rude approximation to describe individual hadron
properties. The problem of proving factorization inside the $LEET$
therefore is not a well posed one.

\noindent
$ii)$
The $LEET$ is capable of describing {\it seminclusive} reactions,
such as (\ref{eq8}),
in which high energy partons, evolving into hadronic jets,
are replaced by effective quarks.
For the latter processes factorization can be proved in a
non perturbative and gauge invariant way with the theory 
of Wilson loops. 
The proof
is a formalization of the Bjorken color dipole argument
presented in the introduction. 

\noindent
$iii)$
Exclusive decays can be studied in the framework of a more
complicated effective theory than the $LEET$, 
which accounts for transverse momentum dynamics in first
approximation. 

\noindent
$iiii)$
It is not clear at present if Bjorken idea can be formalized
for exclusive decays in the framework of the new effective theory. 
That is because transverse momentum terms produce
a diffusion of Wilson lines off the classical quark trajectory.
The only clue is that the
form of the propagator (\ref{eq18}) shows large interference
effects in the quark wave function away from the
classical trajectory.
That gives perhaps some indication
of a small effective color dipole interaction
of the system of light quarks in the decay (\ref{eq1}).

\end{document}